\documentclass[12pt]{article}
\usepackage{graphicx}
\newcommand\pubnumber{DPF2013-103}
\newcommand\pubdate{\today}

\def\support{\footnote{
G.~Aglieri Rinella, F.~Ambrosino, B.~Angelucci, A.~Antonelli, G.~Anzivino, 
R.~Arcidiacono, I.~Azhinenko, S.~Balev, A.~Biagioni, C.~Biino
% ...
}}

\def\Title#1{\begin{center} {\Large #1 } \end{center}}
\def\Author#1{\begin{center}{ \sc #1} \end{center}}
\def\Address#1{\begin{center}{ \it #1} \end{center}}
\newcommand\pubblock{\rightline{\begin{tabular}{l} \pubnumber\\
         \pubdate  \end{tabular}}}
\newenvironment{Abstract}{\begin{quotation}  }{\end{quotation}}
\newenvironment{Presented}{\begin{quotation} \begin{center} 
             PRESENTED AT\end{center}\bigskip 
      \begin{center}\begin{large}}{\end{large}\end{center} \end{quotation}}

\begin{document}
\begin{titlepage}
\pubblock

\vfill
\Title{Searches for rare and forbidden kaon decays at the NA62 experiment 
at CERN}
\vfill
\Author{Matthew Moulson, for the NA62 Collaboration\support}
\Address{Laboratori Nazionali di Frascati dell'INFN, 00044 Frascati, Italy}
\vfill
\begin{Abstract}
The decay $K^+ \to \pi^+\nu\bar{\nu}$ is highly suppressed in the 
Standard Model (SM), while its rate can be predicted with minimal
theoretical uncertainty. The branching ratio (BR) for this decay is thus a 
sensitive probe of the flavor sector of the SM; its measurement, however, 
is a significant experimental challenge. The primary goal of the NA62 
experiment is to measure ${\rm BR}(K^+ \to \pi^+\nu\bar{\nu})$ with 
$\sim$10\% precision. This will require the observation of $10^{13}$ $K^+$ 
decays in the experiment's fiducial volume, as well as the use of 
high-performance systems for precision tracking, particle identification, 
and photon vetoing. These aspects of the experiment will also allow NA62 
to carry out a rich program of searches for lepton flavor and/or number
violating $K^+$ decays. Part of the experimental apparatus was commissioned
during a technical run in 2012; installation continues and data taking 
is expected to begin in late 2014. The physics prospects and the status 
of the NA62 experiment are reviewed.
\end{Abstract}
\vfill
\begin{Presented}
DPF 2013\\
The Meeting of the American Physical Society\\
Division of Particles and Fields\\
Santa Cruz, California, August 13--17, 2013\\
\end{Presented}
\vfill
\end{titlepage}
\def\thefootnote{\fnsymbol{footnote}}
\setcounter{footnote}{0}

\section{Introduction}

\begin{figure}[ht]
\centering
\includegraphics[width=80mm]{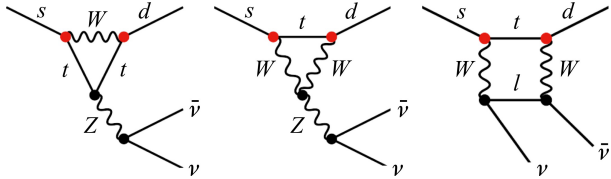}
\caption{Diagrams contributing to the process $K\to\pi\nu\bar{\nu}$.} 
\label{fig:fcnc}
\end{figure}

The $K\to\pi\nu\bar{\nu}$ decays are flavor-changing neutral current (FCNC)
processes that probe the $s\to d\nu\bar{\nu}$ transition via the 
$Z$-penguin and box diagrams shown in Figure~\ref{fig:fcnc}. They are 
highly GIM suppressed and their Standard Model (SM) rates are very small.
For several reasons, the SM calculation for their branching ratios
(BRs) is particularly clean (see \cite{Cirigliano:2011ny} for a recent review):
\begin{itemize}
\item The loop amplitudes are dominated by the top-quark contributions.
The neutral decay violates $CP$; its amplitude involves the top-quark 
contribution only. Small corrections to the amplitudes from the lighter
quarks come into play for the charged channel. 
\item The hadronic matrix element for these decays can be obtained from the
precise experimental measurement of the $K_{e3}$ rate.
\item There are no long-distance contributions from processes with
intermediate photons. 
\end{itemize}
In the SM,
${\rm BR}(K_L\to\pi^0\nu\bar{\nu}) = 2.43(0.39)(0.06)\times10^{-11}$ and
${\rm BR}(K^+\to\pi^+\nu\bar{\nu}) = 7.81(0.75)(0.29)\times10^{-11}$
\cite{Brod:2010hi}.
The uncertainties listed first derive from the input parameters.
The smaller uncertainties listed second demonstrate the size of the
intrinsic theoretical uncertainties. Because of the corrections from 
lighter-quark contributions, these are slightly larger for the charged
channel. 

Because the SM rates are small and predicted very precisely,
the BRs for these decays are sensitive probes for new physics. 
In evaluating the rates for the different FCNC kaon decays, the 
different terms of the operator product expansion are differently
sensitive to modifications from a given new-physics scenario. If
${\rm BR}(K_L\to\pi^0\nu\bar{\nu})$ and ${\rm BR}(K^+\to\pi^+\nu\bar{\nu})$
are ultimately both measured, and one or both BRs is found to differ from
its SM value, it may be possible to characterize the physical mechanism 
responsible \cite{Straub:2010ih}, e.g., a mechanism with minimal flavor 
violation \cite{Hurth:2008jc}, manifestations of supersymmetry 
\cite{Isidori:2006qy}, a fourth generation of fermions \cite{Buras:2010cp}, 
Higgs compositeness as in the littlest Higgs model \cite{Blanke:2009am},
or an extra-dimensional mechanism such as in the 
Randall-Sundrum model \cite{Blanke:2008yr}. 

The decay ${\rm BR}(K_L\to\pi^0\nu\bar{\nu})$ has never been measured 
(the KOTO experiment at J-PARC \cite{Yamanaka:2012yma} has a good chance 
of observing it). ${\rm BR}(K^+\to\pi^+\nu\bar{\nu})$ was measured by 
Brookhaven experiment E787 and its successor, E949. The combined result
from the two generations of the experiment, obtained with seven candidate 
events, is ${\rm BR}(K^+\to\pi^+\nu\bar{\nu}) = 
1.73^{+1.15}_{-1.05}\times10^{-10}$ \cite{Artamonov:2009sz}. 
The purpose of the NA62 experiment at the CERN SPS is to measure 
${\rm BR}(K^+\to\pi^+\nu\bar{\nu})$ with a precision of about 10\% in 
two-years' worth of data taking. 
Observation of $\sim$100 signal events will require a sample of $10^{13}$ 
$K^+$ decays within the geometrical acceptance of the experiment, 
for which the signal detection efficiency must be at least 10\%. 
Then, for a measurement with 10\% precision, the background level must be
kept down to no more than about 10\% of signal. This implies an overall
background rejection factor of $10^{12}$. The residual background level
must also be determined to within about 10\%.

\section{The NA62 experiment}
\begin{figure}[ht]
\centering
\includegraphics[angle=270,width=0.80\textwidth]{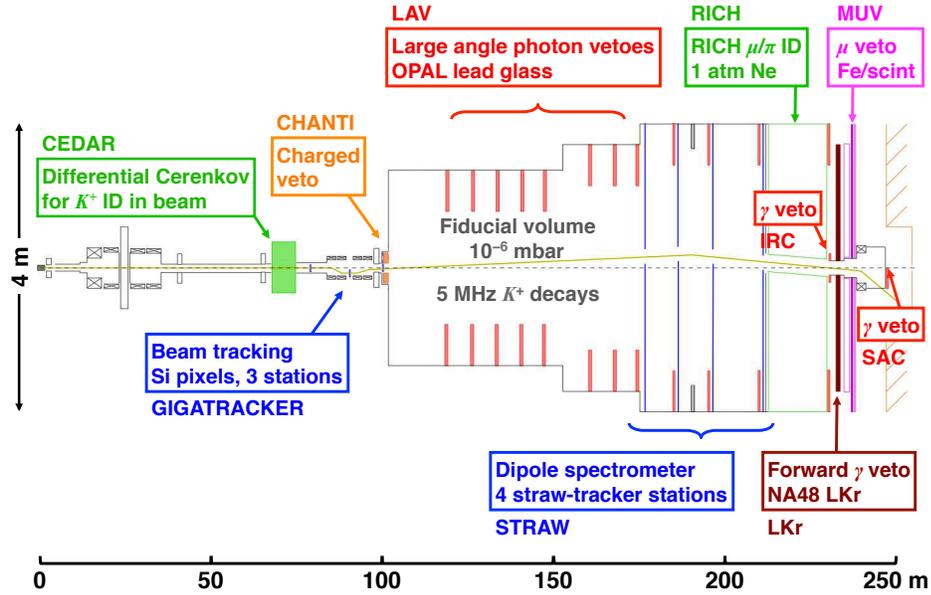}
\caption{Schematic diagram of the NA62 experiment.}
\label{fig:na62}
\end{figure}

The experimental signature is a $K^+$ coming into the experiment
and decaying to a $\pi^+$, with no other particles present. The first 
line of defense against abundant decays such as $K\to\mu\nu$ and 
$K\to\pi\pi^0$ (together representing about 84\% of the total $K^+$ width) 
is to precisely reconstruct the missing mass of the primary and secondary tracks
and reject events with $M_{\rm miss}^2 \approx 0$ or $M_{\rm miss}^2 \approx 
m_{\pi^0}^2$, assuming the secondary is a $\mu^+$ or a $\pi^+$, respectively. 
However, the rejection power from kinematics alone is at best $10^{4}$, and in 
any case, about 8\% of $K^+$ decays (e.g., $K_{e3}$, $K_{\mu3}$) do not have 
closed kinematics. The remainder of the experiment's rejection power must 
come from redundant particle identification systems and hermetic, 
highly-efficient photon veto detectors. The NA62 apparatus \cite{NA62:2010xxx},
schematically illustrated in Figure~\ref{fig:na62}, was designed around 
these principles, which we now consider in turn.

\paragraph{Beamline and decay volume}
The experiment makes use of a 400-GeV primary proton beam from the SPS with 
$3\times10^{12}$ protons per pulse and a duty factor of about 0.3. This is 
collided on a beryllium target at zero angle to produce the 75-GeV $\pm1\%$ 
unseparated positive secondary beam used by the experiment. This beam consists
of about 525~MHz of $\pi^+$, 170~MHz of $p$, and 45~MHz of $K^+$, for a total
rate of 750~MHz. The beamline opens into the vacuum tank about 100~m downstream
of the target. The vacuum tank is about 110~m long and fully encloses the 
four tracking stations of the magnetic spectrometer; the pressure inside 
is kept at a level of $10^{-6}$ mbar. The fiducial volume occupies the
first 60~m of the vacuum tank (upstream of the spectrometer). About 
10\% of the $K^+$'s entering the experiment decay in the fiducial volume,
corresponding to 4.5~MHz of $K^+$ decays.

\paragraph{High-rate, precision tracking}
In order to obtain the full kinematic rejection factor of $10^{4}$ 
for two-body decays, both the beam particle and the decay secondary 
must be accurately tracked.

The beam spectrometer \cite{Fiorini:2013xya} consists of three hybrid 
silicon pixel tracking detectors installed in an achromat in the beam line. 
Each detector consists of a 200-$\mu$m-thick monolithic sensor and 10 
bump-bonded, 100-$\mu$m-thick readout ASICs. The pixel size is 
$300\times300~\mu$m$^2$, giving a momentum resolution 
$\sigma_p/p \sim 0.2\%$ and an angular resolution 
$\sigma_\theta = 16$~$\mu$rad. This beam-tracking system is referred to 
as the Gigatracker, because it will track the individual particles in 
the 750-MHz secondary beam.   

The magnetic spectrometer for the secondary particles consists of four straw
chambers operated inside the vacuum tank \cite{Danielson:2010fta}.
Each chamber has 16 layers of straw tubes arranged in 4 views.
The straws are made from metalized, 36-$\mu$m-thick mylar ultrasonically
welded along the seam. They are just under 10~mm in diameter and are 
2.1~m long. With a 70\% Ar--30\% ${\rm CO_2}$ gas mixture, the point 
resolution on a single view is $\sigma_x \leq 130~\mu$m.
Considering that each chamber is only $0.45\,X_0$ thick, with the 
spectometer magnet providing a $p_\perp$ kick of 270~MeV,
the momentum resolution for tracks is $\sigma_p/p = 0.32\% \oplus 0.008\%\,p$.

\paragraph{Redundant particle identification}
The principal PID challenge for single tracks is to reject $K\to\mu\nu$ 
decays with an inefficiency of less than $10^{-7}$ after the application
of kinematic cuts. The bulk of NA62's $\pi/\mu$ separation capability is 
provided by the downstream muon vetoes (MUV). There are three MUV systems. 
MUVs 1 and 2 are iron/scintillator hadron calorimeters. These are used 
mainly for offline $\mu$ identification and provide a rejection factor 
of $10^5$. MUV 3 is highly segmented and provides fast $\mu$ identification 
for triggering. It can veto $\mu$'s online at a rate of 10~MHz with a time 
resolution $\sigma_t < 1$~ns.  

An additional two orders of magnitude in $\pi/\mu$ separation are provided 
by a large (3.7-m-diameter by 18-m-long) ring-imaging Cerenkov counter (RICH)
\cite{Bucci:2010zz} filled with neon gas at 1~atm ($p_{\rm thresh}$ = 12~GeV 
for $\pi$). In addition to providing good $\pi/\mu$ discrimination over 
the entire fiducial momentum interval ($15 < p < 35$~GeV), the RICH measures 
the $\pi$ crossing time with a resolution of $100$~ps and contributes to 
the level-0 trigger. 

\paragraph{Beam timing and PID}
Considering that the rates of primary and secondary tracks in the experiment 
are respectively about 750~MHz and 10~MHz, accurately matching the correct
secondary track to the correct primary is a basic challenge for the experiment.
Due to the effectively incorrect reconstruction of the primary, for mismatched
events the missing mass resolution is worsened by a factor of three. Precise
timing of the secondary can be obtained from the RICH ($\sigma_t \sim 100$~ps),
while for the primary, the Gigatracker provides $\sigma_t \sim 150$~ps. 
Cerenkov identification of the kaons in the beam both provides a precise,
redundant measurement of the beam particle's timing and reduces the effective
beam rate from 750~MHz to 45~MHz, hence reducing the mismatch probability. 

Such identification is provided by the CEDAR/KTAG, a differential Cerenkov 
counter based on the CERN CEDAR-W design \cite{Romano:2011xxx}. One of the
CEDAR-W detectors has been refurbished to run with ${\rm H}_2$ at 3.85~bar 
and outfitted with a new, high-segmentation readout (KTAG). The beam 
identification from the CEDAR/KTAG is fundamental to the suppression of 
background from beam-gas interactions---without it, the vacuum in the decay
tank would have to be kept at the level of $10^{-8}$~mbar.

With the help from the CEDAR/KTAG, the probability of mismatching the
primary and secondary tracks is held below 1\%. Nevertheless, events with
mismatched tracks still account for half of the events not rejected by
kinematics.

\paragraph{Hermetic photon vetoes}
Rejection of photons from $\pi^0$'s is important for the elimination of 
many background channels. The most demanding task is the rejection of
$K^+\to\pi^+\pi^0$ decays. For these decays, requiring the secondary 
$\pi^+$ to have $p < 35$~GeV guarantees that the two photons from the $\pi^0$
have a total energy of 40~GeV. If the missing-mass cuts provide a 
rejection power of $10^4$, the probability for the photon vetoes
to miss both photons must be less than $10^{-8}$.
The photon veto system consists of four separate subdetector systems.
The ring-shaped large-angle photon vetoes (LAVs) are placed at 12 stations 
along the vacuum volume and provide coverage for decay photons with 
$8.5~{\rm mrad}<\theta<50~{\rm mrad}$.
Downstream of the RICH, the NA48 liquid-krypton calorimeter (LKr) vetoes 
forward ($1~{\rm mrad}<\theta<8.5~{\rm mrad}$), high-energy photons.
A ring-shaped shashlyk calorimeter (IRC) about the 
beamline provides coverage for photons with $\theta<1~{\rm mrad}$, while
further downstream, a small-angle shashlyk calorimeter (SAC) 
around which the beam is deflected completes the coverage for 
very-small-angle photons that would otherwise escape via the beam pipe.

In more than 80\% of $K^+\to\pi^+\pi^0$ events, both photons from the $\pi^0$
arrive at the LKr. In most of the rest of the events, one photon is on the 
LKr and one is in the LAVs. For kinematic reasons, the energies of the two 
photons are anticorrelated: in events with a photon in the LAVs, the energy
of the photon in the LKr tends to be quite high. Given these considerations,
in order to achieve the required $\pi^0$ rejection performance, the LAVs must
have a maximum inefficiency of $10^{-4}$ for photons with $E>200$~MeV, while 
the LKr must have a maximum inefficiency of $10^{-3}$ for photons with 
$E>1$~GeV and $10^{-5}$ for photons with $E>10$~GeV.
The LAV detectors consist of rings of lead-glass blocks salvaged from the 
OPAL electromagnetic calorimeter barrel \cite{Ambrosino:2011xxx}.
The detection efficiency of these blocks for 200~MeV electrons was 
measured at the Frascati BTF and found to be about $(1\pm1)\times10^{-4}$. 
The LKr is a quasi-homogeneous ionization calorimeter of depth $27\,X_0$
and with a transverse segmentation of $2\times2$~cm$^2$ \cite{Fanti:2007vi}. 
In NA48, $K\to\pi\pi^0$ and $e^-$ bremsstrahlung events were used to
demonstrate that the inefficiency of the LKr for detection of photons
with $E>10$~GeV is less than $8\times10^{-6}$.  

\paragraph{Trigger and data acquisition}
The experiment makes use of an integrated trigger and data acqusition system 
with three trigger levels. The lowest level, level 0, is implemented 
directly in the digital readout card for each detector subsystem.
The detector hits are resolved into quantities such as the number of 
quadrants of the trigger hodoscope hit, the number of LKr clusters of energy
greater than a given threshold, or the number of hits in MUV 3.
These quantities can then be used in trigger logic to decide which events
will be read out for level 1. Level 0 will process about 
10 MHz of ``primitive'' detector hits; about 1 MHz of events will be read out
for level 1.
The level 1 trigger is implemented in software running on dedicated PCs for 
each detector. It is the first asynchronous trigger level and will reduce
the rate of events seen by level 2 by an order of magnitude. The level 2
trigger is implemented in the event builder running on the acquisition PC
farm; it is the first trigger level at which the configurations of entire
events are used. The O(100~kHz) of events input to level 2 are reduced to 
a few kHz of events ultimately written to disk. 

\paragraph{Expected performance}
Based on the above considerations, the event selection criteria can be listed:
\begin{itemize}
\item One track with $15<p_\pi<35$~GeV and $\pi$ identification in the RICH.
\item No $\gamma$'s in the LAVs, LKr, IRC, or SAC.
\item No $\mu$ hits in the MUVs.
\item One beam particle in the Gigatracker with $K$ identification by the
CEDAR.
\item $z_{\rm rec}$, the vertex between primary and secondary tracks, inside 
the 60-m fiducial volume.
\end{itemize}
Simulations then indicate that the acceptance for signal events is a little 
more than 10\%, corresponding to about 45 signal events accepted per year
of data taking. The $\pi^+\pi^0$ background is estimated to be about 10\%
while the $\mu\nu$ background is around 3\%. Including backgrounds from all 
other channels, the total background is under 20\%.    

\section{Other rare kaon and pion decays at NA62}

The measurement of ${\rm BR}(K^+\to\pi^+\nu\bar{\nu})$ will require a sample
of $10^{13}$ $K^+$ decays in NA62's fiducial volume. These will be accompanied
by $2\times10^{12}$ $\pi^0$ decays from $K\to\pi\pi^0$ (BR = 21\%). Studies
of the prospects for searches for lepton-flavor (LF) or -number (LN) 
violating and other forbidden decays with NA62 are underway.
Preliminary estimates of the single-event sensitivties (defined as the
reciprocal of the product of the number of accepted decays) give results
at the level of $10^{-12}$ for $K^+$ decays to states such as 
$\pi^+\mu^\pm e^\mp$ (LFV), $\pi^-\mu^+e^+$ (LFNV), and $\pi^-e^+e^+$ or   
$\pi^-\mu^+\mu^+$ (LNV); and at the level of $10^{-11}$ for $\pi^0$ decays
to $\mu^\pm e^\mp$ \cite{Moulson:2013oga}.

As a case in point, consider the decay $K^+\to\pi^-\mu^+\mu^+$. This decay 
violates the conservation of lepton number. In analogy to the case 
of neutrinoless nuclear double beta decay, its observation would imply that
the virtual neutrino exchanged between the $\mu^+$'s annhilates 
itself---the neutrino must have a Majorana component. The most stringent
limit on BR($K^+\to\pi^-\mu^+\mu^+$) is from NA48/2 \cite{Batley:2011zz}, 
and NA62's prospects for improving on this limit can be extrapolated from 
the NA48/2 experience. In a sample of $2\times10^{11}$ $K^\pm$ decays, NA48/2
had 52 candidate events selected as $\pi^\mp\mu^\pm\mu^\pm$ for which 
$M(\pi\mu\mu) \sim m_K$. This was in excellent agreement with the Monte Carlo
background estimate and gave the published result, 
${\rm BR}(K^\pm\to\pi^\mp\mu^\pm\mu^\pm) < 1.1\times10^{-9}$ (90\%~CL).
However, subsequent studies showed that the background consisted entirely
of $K^\pm\to\pi^\mp\pi^\pm\pi^\pm$ events with two $\pi\to\mu$ decays, of which
at least one was downstream of the spectrometer magnet and therefore poorly
reconstructed. As it turns out, the increased $p_\perp$ kick of the NA62 
magnet, together with the better invariant mass resolution of the straw-tube 
spectrometer, can eliminate this background altogether. It is then quite 
possible for NA62 to push the limit on this BR all the way down to its 
single-event sensitivity of order $10^{-12}$.

Besides the LFV $\pi^0$ decays, there are a number of rare or forbidden 
$\pi^0$ decays to which NA62 has potential sensitivity, including 
$\pi^0\to3\gamma$, $\pi^0\to4\gamma$, and $\pi^0\to e^+e^-e^+e^-$
 \cite{Moulson:2013oga}.

One interesting prospect is to examine $e^+e^-\gamma$ final states of
$\pi^0$ decays for evidence for a new, light vector gauge boson with weak
couplings to charged SM fermions, a so-called $U$ boson, or ``dark photon''. 
A hypothetical $U$ boson could mediate the interactions of dark-matter
constituents, as such providing explanations for various unexpected 
astrophysical observations and the results of certain dark-matter searches, 
and could also explain the $>3\sigma$ discrepancy between the measured 
and predicted values for the muon anomaly, $a_\mu$ 
(see e.g. \cite{Pospelov:2008jk,Pospelov:2008zw}).
A $U$ boson with a mass of less than $m_{\pi^0}/2$ might be directly 
observable in $\pi^0\to U\gamma$ decays with $U\to e^+e^-$.
Using an appropriate trigger, NA62 may collect
$\sim$$10^8$ $\pi^0\to e^+e^-\gamma$ decays per year.
Moreover, NA62 has good invariant-mass resolution for the $ee$ pair---about 
1~MeV even before any attempt at kinematic fitting. Thus, NA62 should be
quite competitive in this search.

Another possibility is to search for the 
invisible decay of the $\pi^0$. The least exotic decay to an invisible
final state is $\pi^0\to\nu\bar{\nu}$. This is forbidden by angular-momentum
conservation if neutrinos are massless; for a massive neutrino $\nu$ of 
a given flavor and mass $m_\nu < m_{\pi^0}/2$ with standard coupling to the 
$Z$, the calculation of the decay rate is straightforward. 
The experimental signature $\pi^0\to{\rm invisible}$ could also arise 
from $\pi^0$ decays to other weakly interacting neutral states. 
Experimentally, the process $K^+\to\pi^+\pi^0$ with $\pi^0\to{\rm invisible}$
is very similar to $K^+\to\pi^+\nu\bar{\nu}$, with the important difference
that in the former case, the $\pi^+$ is monochromatic in the rest frame of 
the $K^+$. This means that there is no help from kinematics in identifying
$K^+\to\pi^+\pi^0$, $\pi^0\to\gamma\gamma$ with two lost photons---the limit on
${\rm BR}(\pi^0\to{\rm invisible})$ essentially depends on the performance 
of the photon vetoes. With stringent track-quality cuts for the $\pi^+$ and
additional cuts in the $(p_{\pi^+}, \theta_{\pi^+})$ plane to deselect 
events with low-energy, large-angle photons, the $\pi^0$ rejection can be 
increased by perhaps a factor of ten with respect to the NA62 baseline
rejection of $10^{-8}$. Then, NA62 would have the potential to set a limit
on ${\rm BR}(\pi^0\to{\rm invisible})$ of $\sim$$10^{-9}$, which is about 100
times better than present limits.  

\section{Outlook}

As of October 2013, the CEDAR/KTAG, almost all of the LAV system, the new LKr 
readout, and the SAC are installed or under installation. The remainder
of the detectors are under construction. The experiment will be ready to
take data in the fall of 2014. A first period of data taking during 
the months of November and December is expected to net the first 
10\% of the NA62 data set. The remainder of the data will be collected 
in long runs in 2015 and 2016. Collection of the full data set will
permit the measurement of ${\rm BR}(K^+\to\pi^+\nu\bar{\nu})$ to within
about 10\%, which should help to shed light on the flavor structure of 
any new physics discovered at the LHC, or which may provide evidence for
new physics even in the absence of such discoveries. NA62 is also well
adapted to search for other rare decays of the $K^+$ and $\pi^0$, with
single-event BR sensitivity at the level of $10^{-12}$ for lepton-flavor
or -number violating decays and competitive prospects in related searches.


\begin{thebibliography}{99}

\bibitem{Cirigliano:2011ny} 
  V.~Cirigliano, G.~Ecker, H.~Neufeld, A.~Pich and J.~Portoles,
  Rev.\ Mod.\ Phys.\  {\bf 84}, 399 (2012)
  [arXiv:1107.6001 [hep-ph]].

\bibitem{Brod:2010hi} 
  J.~Brod, M.~Gorbahn and E.~Stamou,
  Phys.\ Rev.\ D {\bf 83}, 034030 (2011)
  [arXiv:1009.0947 [hep-ph]].

\bibitem{Straub:2010ih} 
  D.~M.~Straub,
  arXiv:1012.3893 [hep-ph].

\bibitem{Hurth:2008jc} 
  T.~Hurth, G.~Isidori, J.~F.~Kamenik and F.~Mescia,
  Nucl.\ Phys.\ B {\bf 808}, 326 (2009)
  [arXiv:0807.5039 [hep-ph]].

\bibitem{Isidori:2006qy} 
  G.~Isidori, F.~Mescia, P.~Paradisi, C.~Smith and S.~Trine,
  JHEP {\bf 0608}, 064 (2006)
  [hep-ph/0604074].

\bibitem{Buras:2010cp} 
  A.~J.~Buras, B.~Duling, T.~Feldmann, T.~Heidsieck and C.~Promberger,
  JHEP {\bf 1009}, 104 (2010)
  [arXiv:1006.5356 [hep-ph]].

\bibitem{Blanke:2009am} 
  M.~Blanke, A.~J.~Buras, B.~Duling, S.~Recksiegel and C.~Tarantino,
  Acta Phys.\ Polon.\ B {\bf 41}, 657 (2010)
  [arXiv:0906.5454 [hep-ph]].

\bibitem{Blanke:2008yr} 
  M.~Blanke, A.~J.~Buras, B.~Duling, K.~Gemmler and S.~Gori,
  JHEP {\bf 0903}, 108 (2009)
  [arXiv:0812.3803 [hep-ph]].

\bibitem{Yamanaka:2012yma}
  T.~Yamanaka [KOTO Collaboration],
  PTEP {\bf 2012} (2012) 02B006.

\bibitem{Artamonov:2009sz} 
  A.~V.~Artamonov {\it et al.}  [BNL-E949 Collaboration],
  Phys.\ Rev.\ D {\bf 79}, 092004 (2009)
  [arXiv:0903.0030 [hep-ex]].

\bibitem{NA62:2010xxx}
  F.~Hahn {\it et al.} (eds.)  [NA62 Collaboration],
  ``NA62: Technical Design Document,''
  NA62-10-07 (2010)

\bibitem{Fiorini:2013xya}
  M.~Fiorini {\it et al.},
  Nucl.\ Instrum.\ Meth.\ A {\bf 718} (2013) 270.

\bibitem{Danielson:2010fta} 
  H.~Danielson [NA62 Collaboration],
  IEEE Nucl.\ Sci.\ Symp.\ Conf.\ Rec.\  {\bf 2010}, 1914 (2010).

\bibitem{Bucci:2010zz} 
  F.~Bucci, G.~Collazuol and A.~Sergi,
  Nucl.\ Instrum.\ Meth.\ A {\bf 623}, 327 (2010).

\bibitem{Romano:2011xxx} 
  A.~Romano [for the NA62 Collaboration],
  Astroparticle, Particle, Space Physics and Detectors for Physics 
  Applications, World Scientific, 895 (2012)

\bibitem{Ambrosino:2011xxx} 
  F.~Ambrosino {\it et al.},
  IEEE Nucl.\ Sci.\ Symp.\ Conf.\ Rec.\  {\bf 2011}, 1159 (2011)
  [arXiv:1111.4075 [physics.ins-det]].

\bibitem{Fanti:2007vi} 
  V.~Fanti {\it et al.}  [NA48 Collaboration],
  Nucl.\ Instrum.\ Meth.\ A {\bf 574}, 433 (2007).

\bibitem{Moulson:2013oga}
  M.~Moulson [for the NA62 Collaboration],
  arXiv:1306.3361 [hep-ex].

\bibitem{Batley:2011zz} 
  J.~R.~Batley {\it et al.}  [NA48/2 Collaboration],
  Phys.\ Lett.\ B {\bf 697}, 107 (2011)
  [arXiv:1011.4817 [hep-ex]].

\bibitem{Pospelov:2008jk} 
  M.~Pospelov, A.~Ritz and M.~B.~Voloshin,
  Phys.\ Rev.\ D {\bf 78}, 115012 (2008)
  [arXiv:0807.3279 [hep-ph]].

\bibitem{Pospelov:2008zw} 
  M.~Pospelov,
  Phys.\ Rev.\ D {\bf 80}, 095002 (2009)
  [arXiv:0811.1030 [hep-ph]].
 
\end{thebibliography}
\end{document}